\documentstyle[aaspp4,psfig,11pt]{article}
\lefthead{J. Yan et al}
\righthead{Wide Field Surveys of Herbig-Haro Objects}

\begin{document}

\title{Wide Field Surveys of Herbig-Haro Objects}

\author{Jun Yan$^{1,2}$, Hongchi Wang$^{1,2}$, Min Wang$^{1,2}$,
Licai Deng$^1$, Ji Yang$^2$ \& Jiansheng Chen$^1$}

\affil{
$^1$Beijing Astronomical Observatory, Academia Sinica, Beijing 100080, PR China\\
$^2$Purple Mountain Observatory, Academia Sinica, Nanjing 210008, PR China
} 
\begin{abstract}

We report our new results from wide-field surveys of Herbig-Haro (HH) objects in the
nearby star forming regions.  The surveys covered approximately 56 
deg$^2$ in Perseus, Taurus, Orion, Monoceros, and other regions.
Using refined techniques,
we discovered in total 68 HH objects, of which 32 were in Perseus,
4 in Taurus, 13 in Orion, 18 in Monoceros, 
and 1 in S287 regions. The newly discovered HH objects demonstrate
a great variety of morphological structures, including 5 jets,  7 arcs,
12 cirri or cirrus groups, 13 patches, and many knots. These objects provide a new
comprehensive database for the study of HH objects in the regions
of recent star formation.

\end{abstract}

\section{Introduction}

Herbig-Haro (HH) objects are small shock-excited nebulae intimately associated
with star forming regions (see reviews by Schwartz 1983 and Raga 1989). Like
high velocity CO molecular outflows and shock-excited near-IR emissions of H$_2$,
HH objects are good tracers of the mass outflow activities of young stellar
objects (YSOs) and can be
used to trace the extremely young class 0 objects embedded in dense molecular
cores -- about 30\% class 0 objects are now known to be
associated with HH objects (Eiroa et al 1994).

\par

The prototype HH objects, HH 1 and 2, were discovered by Herbig (1951) 
and Haro (1952) in their H$_\alpha$ emission star survey. Since then about
300 HH objects have been found by several searching methods, including 
objective-prism Schmidt survey,
narrow-band CCD imaging, near-IR imaging, and other methods (Reipurth 1994).

\par

In this paper, we describe the way of our large-field CCD Schmidt surveys
of HH objects using intermediate band filters, and report the discovery
of a large number of new HH objects in Perseus, Taurus, Orion, Monoceros,
and other star forming 
regions.

\section{Observations and Identification Techniques}

\subsection{Instrumentation and Filters}

The observations of the surveys were carried out at Xinglong Station of
Beijing Astronomical Observatory (BAO) during the winters of 1995-1997.
The telescope used is the BAO f/3 60/90cm Schmidt telescope equipped
with a 2k$\times$2k Aerospace Ford CCD which has a pixel size of 15$\mu$, corresponding
to a resolution of 1.67\arcsec/pixel, and has a total field of view of 58\arcmin. On average,
the seeing at the site is around 2\arcsec.
The filter set used in this program are two BATC\footnote{BATC-Beijing-Arizona-Taiwan-Connetticut Multicolor Sky Survey} intermediate band filters
[BATC09], [BATC10], 
and a narrow band filter [BATC26]. The parameters of
the filters are given in Table~\ref{table1}. As shown in this table,
the [BATC09] filter well covers the strong and characteristic lines of HH objects, 
while the [BATC10] band
has no strong lines of these objects and,
therefore, can be used to represent the continuum.

\begin{table}[h]
\caption{Filter Parameters}\label{table1}
\begin{tabular}{cccl}
\hline
\hline
Filter&Central &Band&Property\\
ID&Wavelength&Width&\\
\hline
{[BATC09]}&6660$\AA$&480$\AA$& [NII], H$_\alpha$, [SII]\\
{[BATC10]}&7050$\AA$&300$\AA$&Continuum\\
{[BATC26]}&6725$\AA$&50$\AA$&[SII] $\lambda\lambda$6717/6731\\
\hline
\end{tabular}
\end{table}

\subsection{Field Selection}   
HH objects are produced by interactions of mass outflows of 
YSOs with the surrounding medium (Schwartz 1975). They are associated with other
tracers of the activities of YSOs, such as molecular
outflows, H$_2$O or OH masers, and they usually occurs in groups. For the
purpose of large field surveys of HH objects in star forming regions,
we have selected 56 fields in the surveys based on the following
selection criteria:

\begin{enumerate}
\item There are known HH objects, molecular outflows, H$_2$O or OH masers
in or near the field;
\item There are IRAS sources of class 0 or class 1, VLA continuum emissions
in the field;
\item There are GGD or RNO objects in the field.
\end{enumerate}

Table 2 gives the log of the surveys.

\begin{table}[p]
\caption{Log of observations}\label{table}
{\tiny
\begin{tabular}{cccccclc}
\hline\hline
Field&\multicolumn{2}{c}{Field Center (1950)}&\multicolumn{5}{c}{Exposure Time and Date}\\
id&R.A.&DEC&[BATC09](min.)&[BATC10](min.)&Run*&[BATC26](min.)&Run*\\
\hline
A01&03:21:30&30:10:00&15&15&A&&\\
A02&03:25:30&30:10:00&15&15&A&60&C\\
A03&03:29:30&30:10:00&15&15&A&60&C\\
A04&03:21:30&31:00:00&15&15&A&&\\
A05&03:25:30&31:00:00&15&15&A&60&C\\
A06&03:29:30&31:00:00&15&15&A&60&C\\
A07&03:41:20&31:25:00&15&15&A&&\\
A08&03:45:20&31:25:00&20&18&A&&\\
A09&03:41:20&32:15:00&15&15&A&&\\
A10&03:45:20&32:15:00&15&15&A&60&C\\
A11&04:24:27&26:00:00&20&20&A&&\\
A12&04:31:17&22:50:00&24&24&A&60&C\\
A13&04:01:41&26:00:00&24&24&A&&\\
A14&04:35:20&25:18:00&15&20&A\\
A15&04:32:00&24:18:00&15&15&A&&\\
A16&04:28:00&24:18:00&15&15&A&&\\
A17&04:15:20&28:00:00&15&15&A&&\\
A18&04:17:00&27:03:00&15&15&A&&\\
A19&04:19:04&19:25:00&15&15&A&&\\
B01&05:48:40&01:55:00&20&20&B&&\\
B02&05:48:40&02:50:00&20&20&B&60&C\\
B03&05:52:20&01:55:00&20&20&B&&\\
B04&05:52:20&02:50:00&20&20&B&60&D\\
B05&06:06:00&-06:20:00&24&24&B&60&C\\
B06&06:09:40&-06:20:00&24&24&B&&\\
B07&04:27:17&22:50:00&24&24&B&&\\
B08&04:35:17&22:50:00&24&24&B&&\\
B09&06:57:30&-04:35:00&24&24&B&60&D\\
B10&06:57:17&-07:42:16&6&24&B\\
B11&03:33:20&30:35:00&20&20&B&&\\
B12&07:02:45&-11:15:00&20&20&B&&\\
B13&06:38:20&10:35:00&20&20&B&60&D\\
B14&03:33:20&31:25:00&20&20&B&&\\
B15&03:37:20&30:35:00&20&20&B&&\\
B16&06:38:20&09:40:00&20&20&B&60&D\\
B17&06:05:46&21:35:00&20&20&B&&\\
B18&03:37:20&31:25:00&20&20&B&&\\
B19&03:37:20&32:15:00&20&20&B&&\\
B20&03:41:20&30:35:00&20&20&B&&\\
B21&06:06:00&18:00:00&20&20&B&&\\
B22&06:09:50&18:00:00&20&20&B&&\\
B23&03:03:32&58:19:00&20&20&B&&\\
B24&03:23:00&58:36:00&20&20&B&&\\
B25&06:11:48&13:51:00&20&20&B&60&D\\
B26&03:45:20&33:05:00&20&20&B&&\\
B27&06:56:46&-03:53:00&20&20&B&&\\
B28&06:45:20&-02:09:30&20&20&B&&\\
B29&06:36:30&08:50:00&20&20&B&&\\
B30&06:31:00&04:10:00&20&20&B&&\\
B31&05:59:00&16:15:00&20&20&B&&\\
B32&05:36:00&35:55:00&20&20&B&&\\
B33&05:34:30&31:58:00&20&20&B&&\\
B34&05:44:00&-00:50:00&30&20&B&&\\
B35&05:44:00&00:04:00&20&20&B&60&D\\
B36&05:44:00&00:58:00&20&20&B&&\\
B37&05:40:20&-01:44:00&20&20&B&60&D\\
\hline

\multicolumn{8}{l}{
	(*) Run code: A-Oct 15,1995$\sim$Dec 20,1995; B-Feb 13,1996$\sim$Mar 18,1996;}\\
\multicolumn{8}{l}{
	C-Dec 6,1996$\sim$Dec 9,1996; D-Mar 7,1997$\sim$Mar 11,1997}\\
\end{tabular}
}
\end{table}

\subsection{Identification Techniques}

Our survey sequence includes a first-step quick survey for HH candidates 
(Runs A and B in Table 2 )
and a narrow-band identification (Runs C and D).
The subsequent observations of this survey
program is in progress, and the details of each target field
will be reported in a later paper (Yan et al. 1997). Here we give the
techniques of picking up new HH objects and some snapshots of our first results. For a target field, 3 or more 
frames in both [BATC09] and [BATC10] bands were taken and combined
so that the cosmic rays and the bad
pixels were eliminated in the resultant images. The two resultant frames in [BATC09] 
and  [BATC10] bands were blinked and compared. Due to their strong emissions in 
H$_\alpha$, [NII] and [SII] lines, HH objects are prominent 
in [BATC09] band but usually invisible in [BATC10] band because of their very low 
continuum emission in this passband. 
HH candidates were picked up from the careful comparison between the two
images. The minimum angular size of detectable condidates is limited
by the pixel size and is about 2\arcsec in our surveys.
In this step we picked up about 150 HH candidates in the 56 fields.
\par

In the first step there is some possibility of contamination by compact HII regions and reflection nebulae. Our second step is to identify the candidates using the narrow-band [BATC26] filter, 
which covers two characteristic lines of HH object, $\lambda\lambda$6717,6731.  We have made this further idendification in 15 fields out of the 56 target fields in table 2. Three frames, each with an exposure time of 20 minutes, were taken; Then the frames were reduced just as in the [BATC09] and [BATC10] bands. We found that above 90 percent of the candidates of HH objects from our first step are identifiable in the [BATC26] band. The images of HH objects are sharper in the [BATC26] frames than in the [BATC09] frames. In contrast, the compact HII regions and reflection nebulae are invisible in the [BATC26] frames.

\begin{figure}[p]
\centerline{\psfig{figure=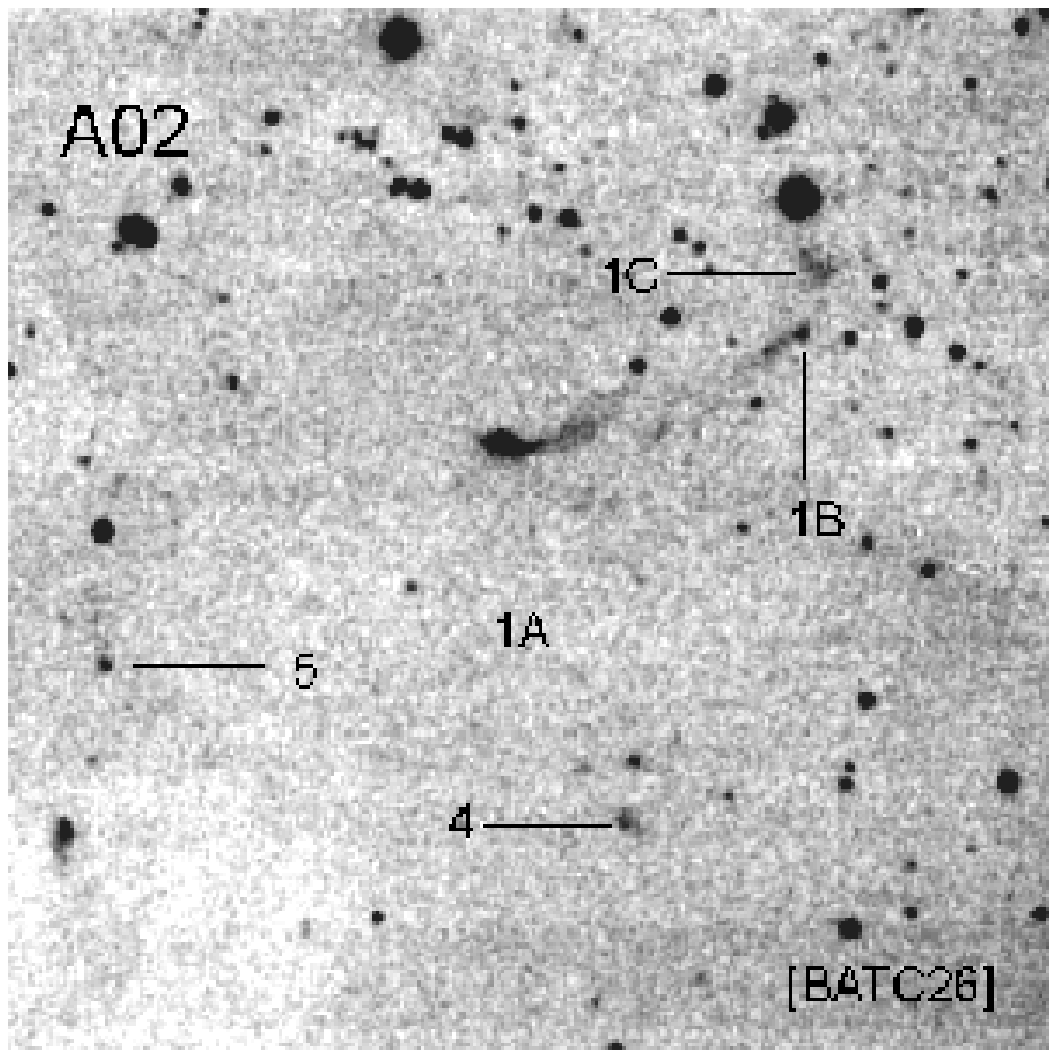,height=5.3cm,width=5.3cm}\hskip 1mm\psfig{figure=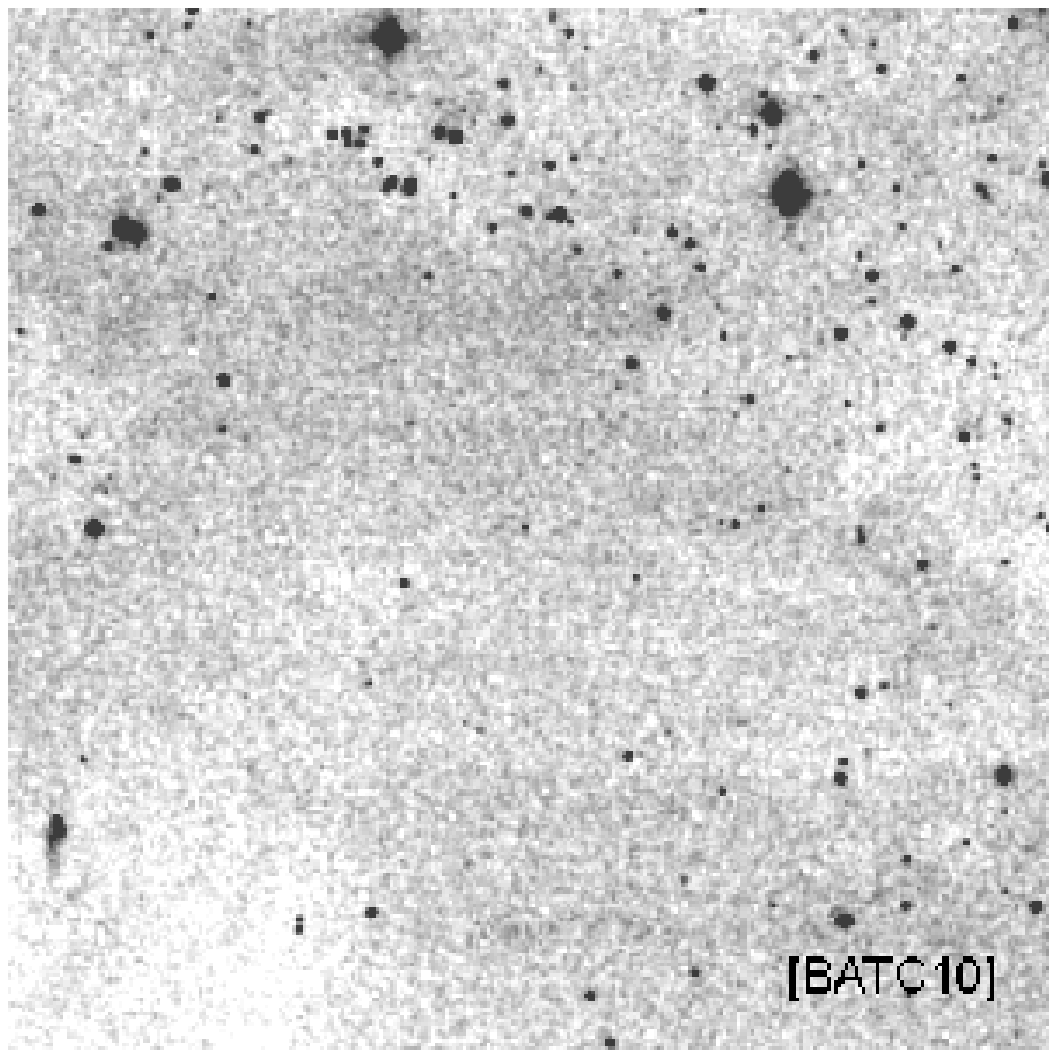,height=5.3cm,width=5.3cm}\hskip 1mm\psfig{figure=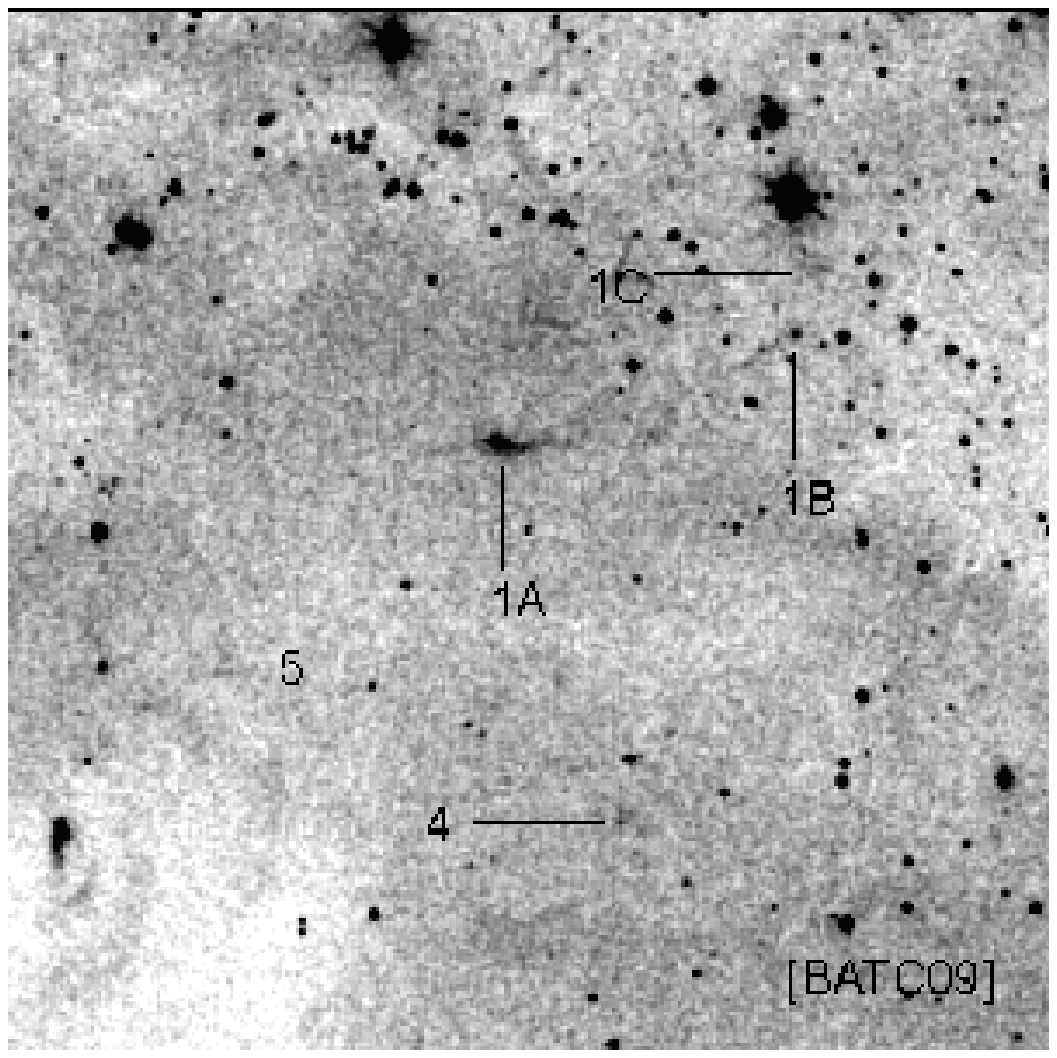,height=5.3cm,width=5.3cm}}
\vskip 1mm
\centerline{\psfig{figure=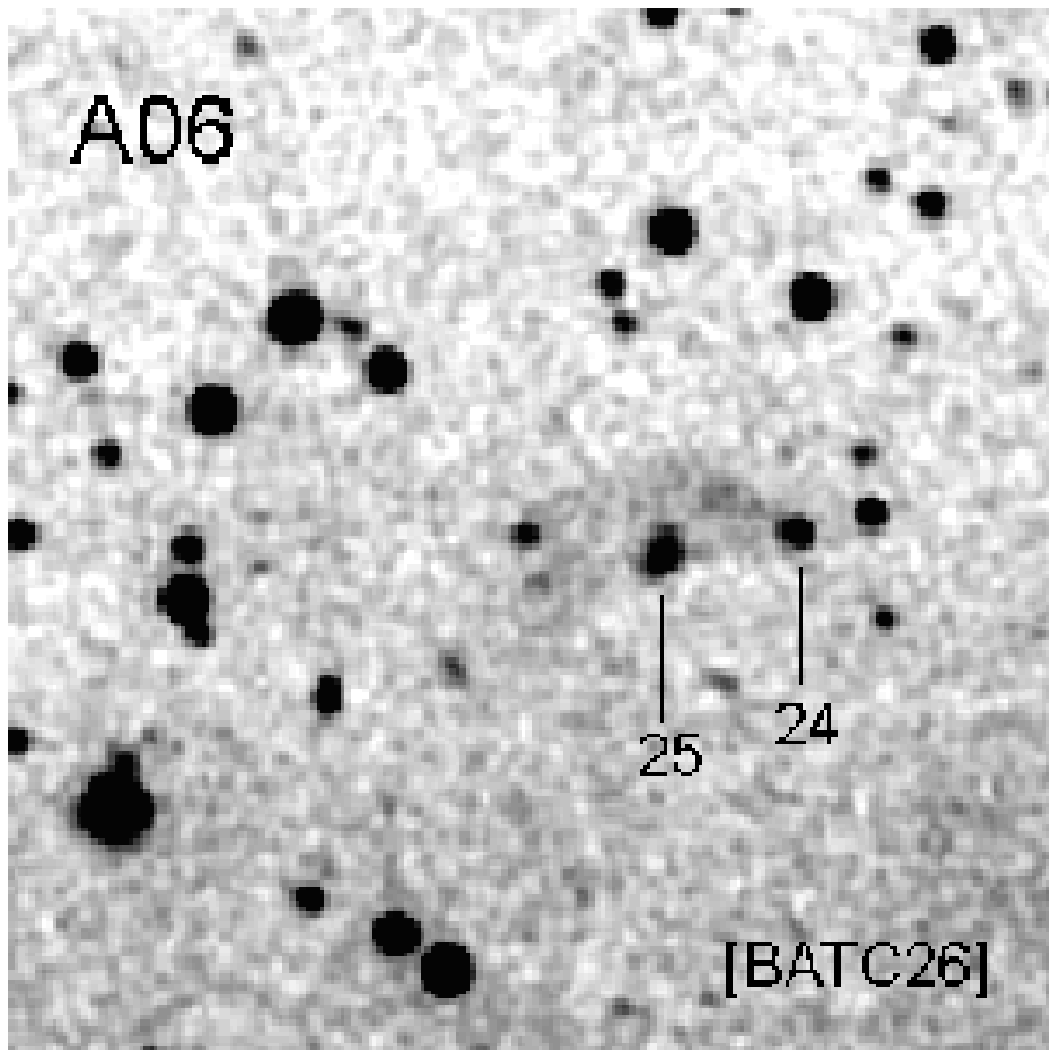,height=5.3cm,width=5.3cm}\hskip 1mm\psfig{figure=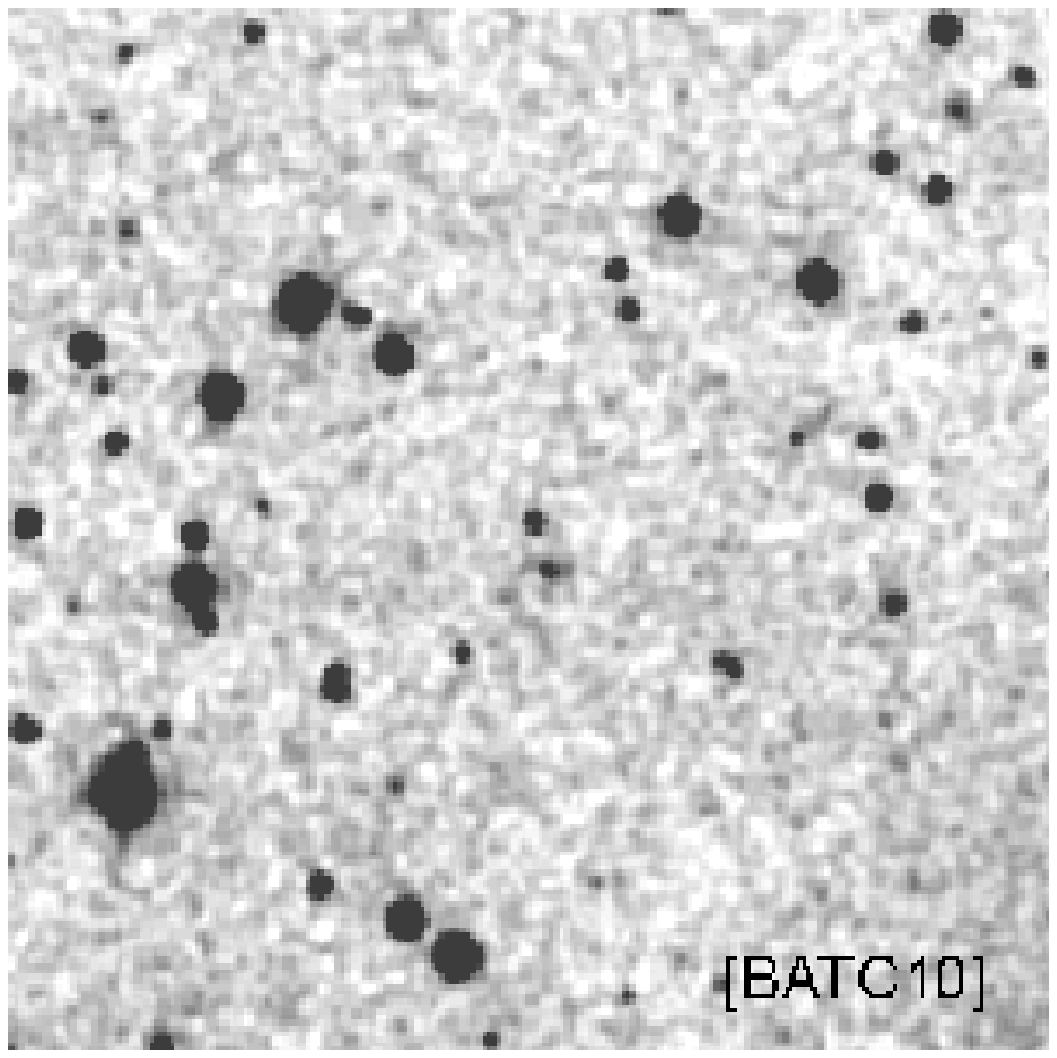,height=5.3cm,width=5.3cm}\hskip 1mm\psfig{figure=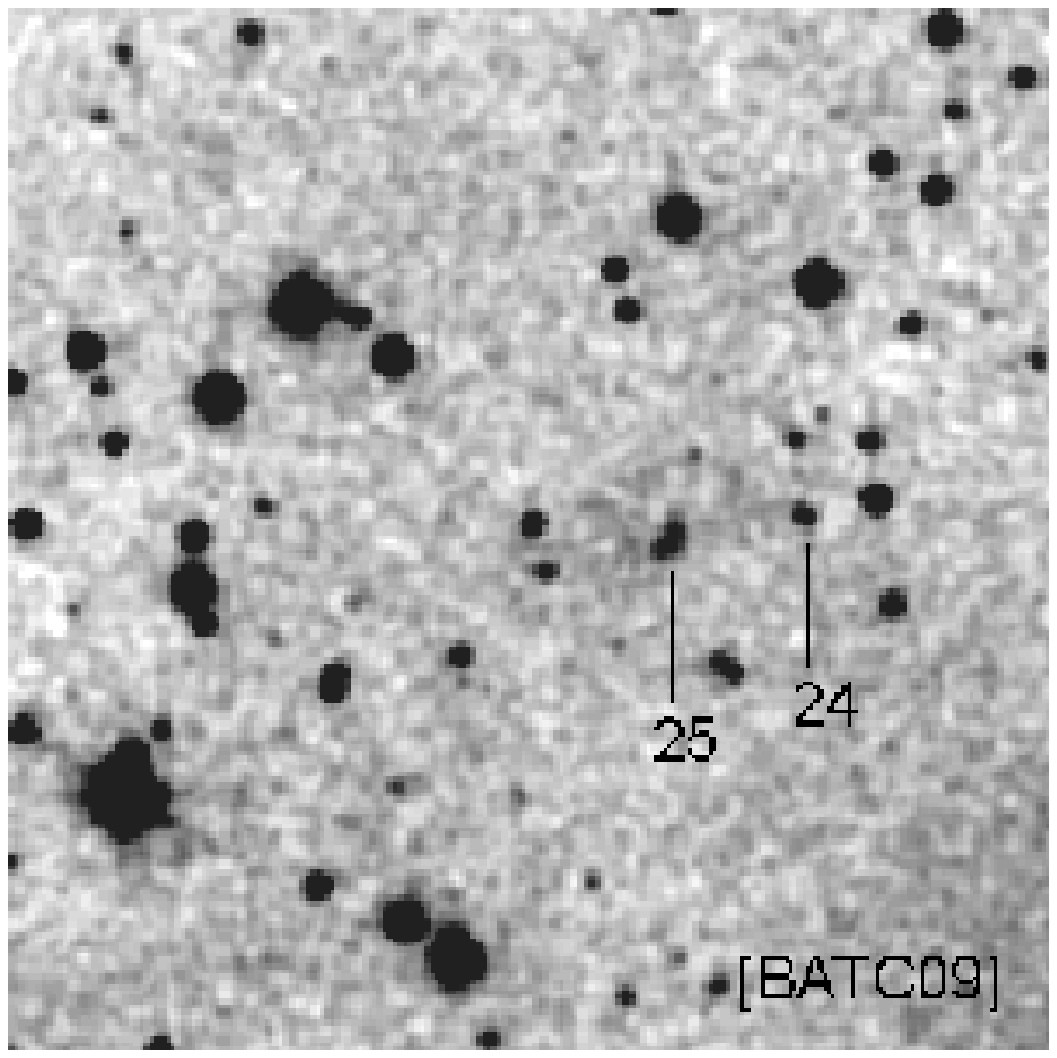,height=5.3cm,width=5.3cm}}
\vskip 1mm
\centerline{\psfig{figure=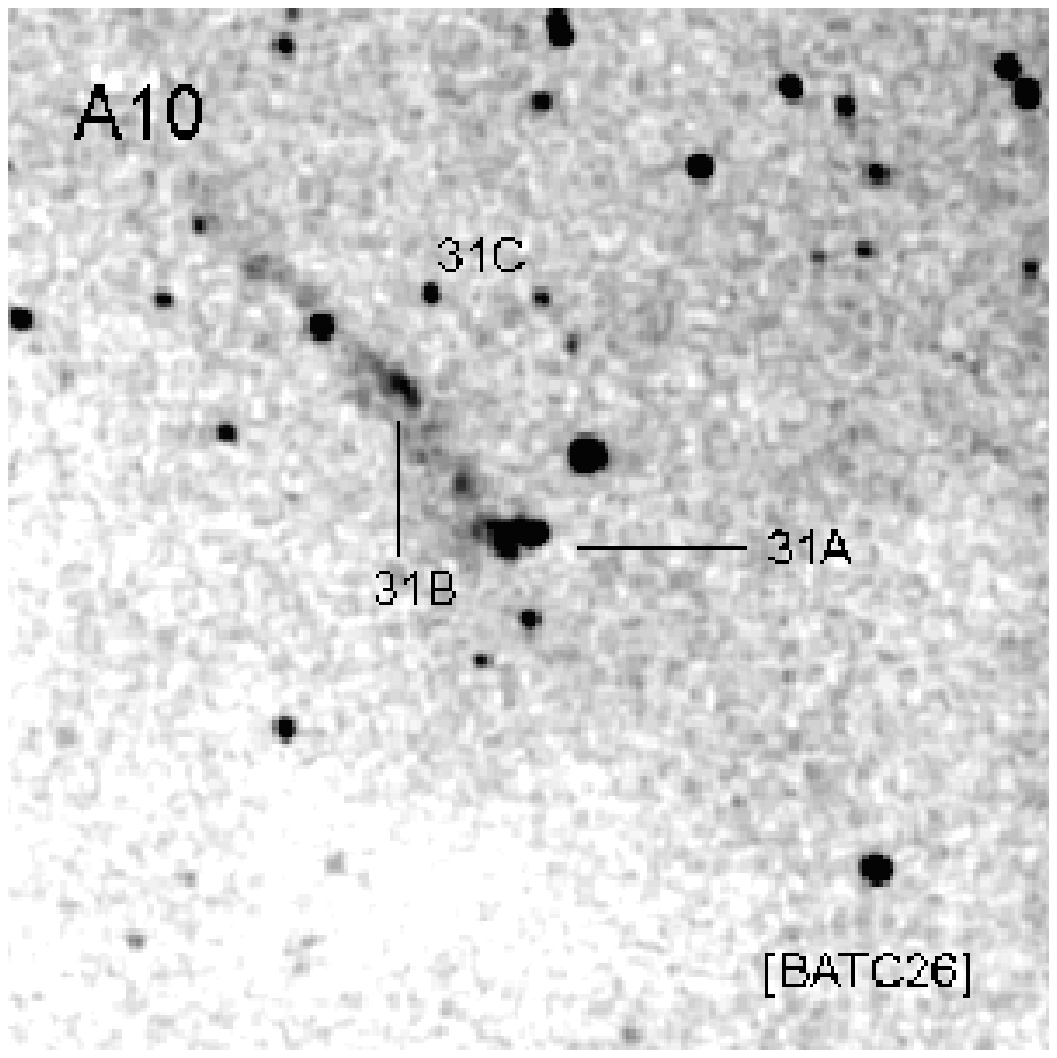,height=5.3cm,width=5.3cm}\hskip 1mm\psfig{figure=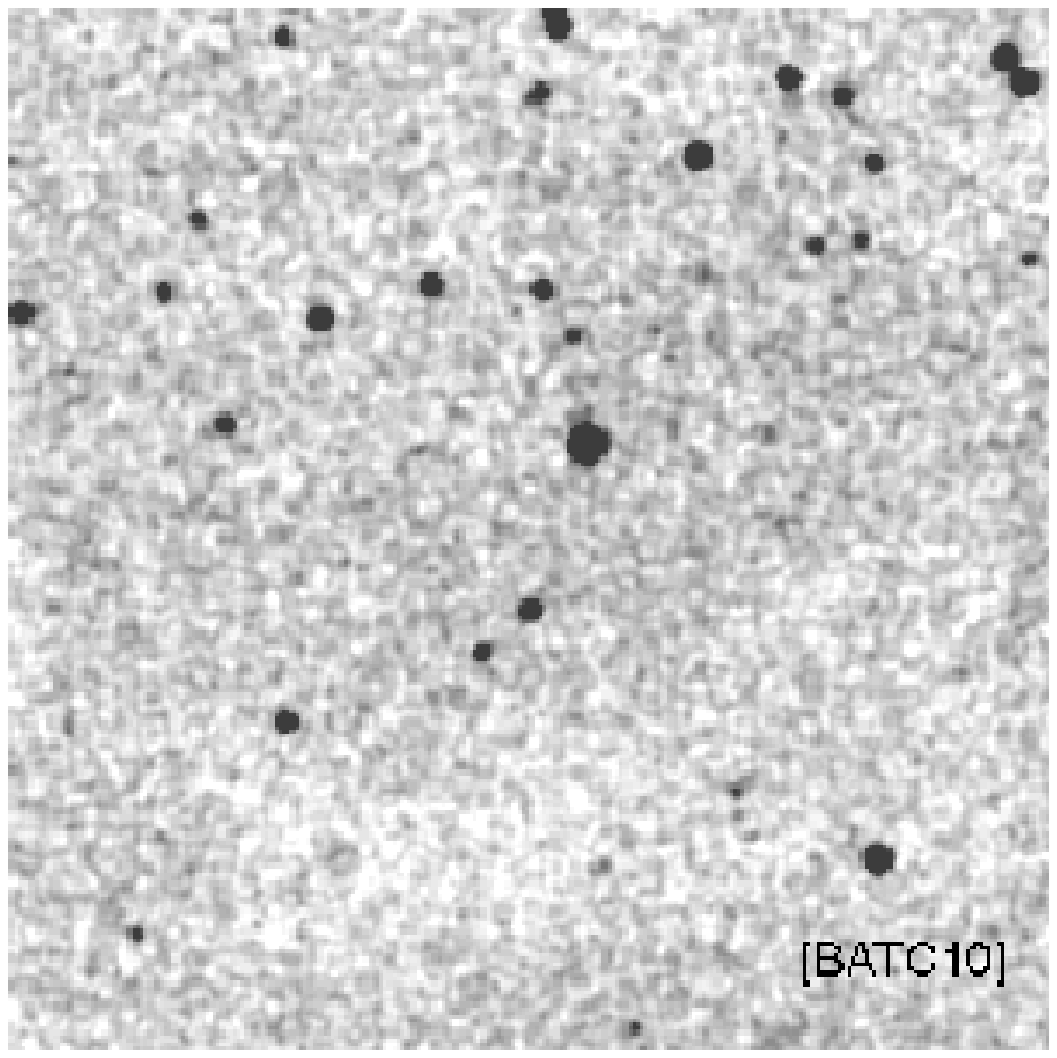,height=5.3cm,width=5.3cm}\hskip 1mm\psfig{figure=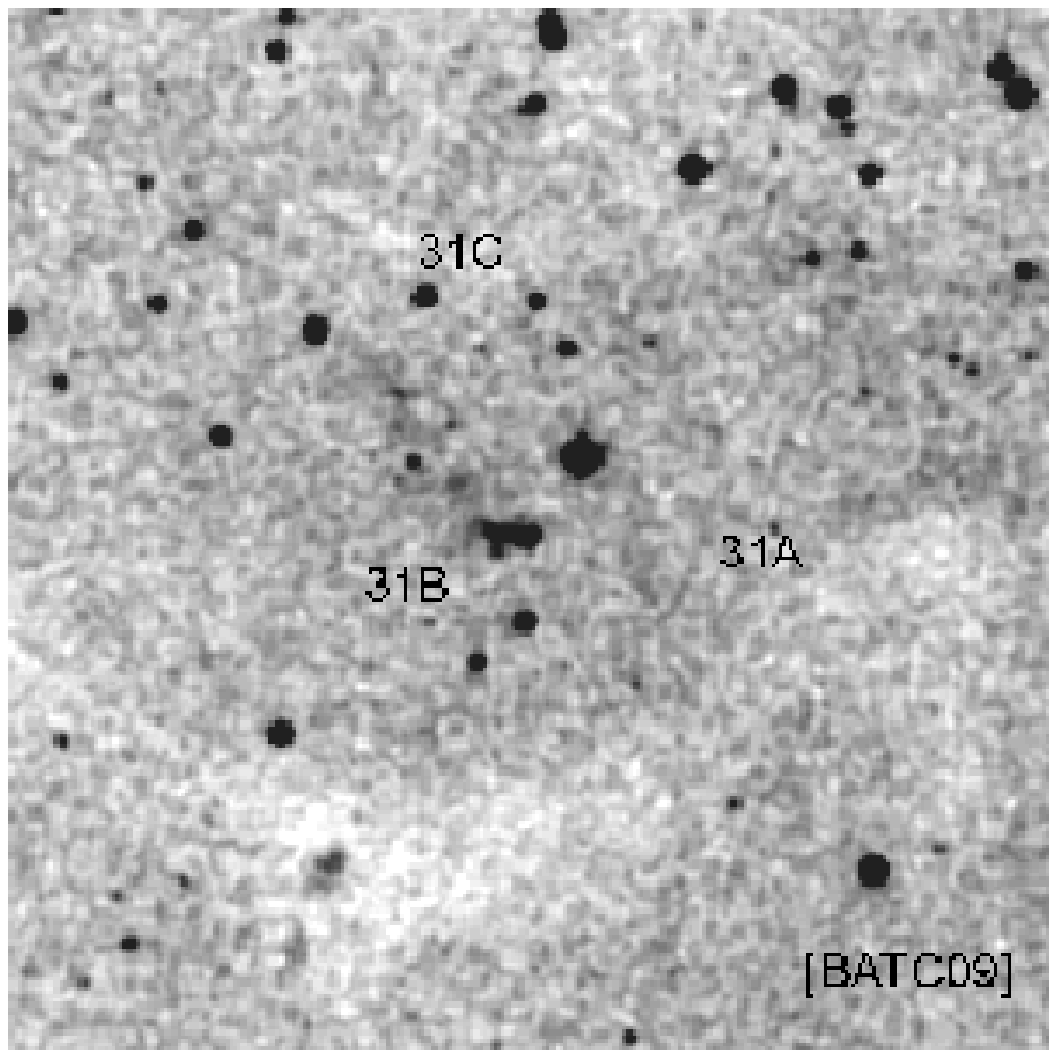,height=5.3cm,width=5.3cm}}
\vskip 1mm
\centerline{\psfig{figure=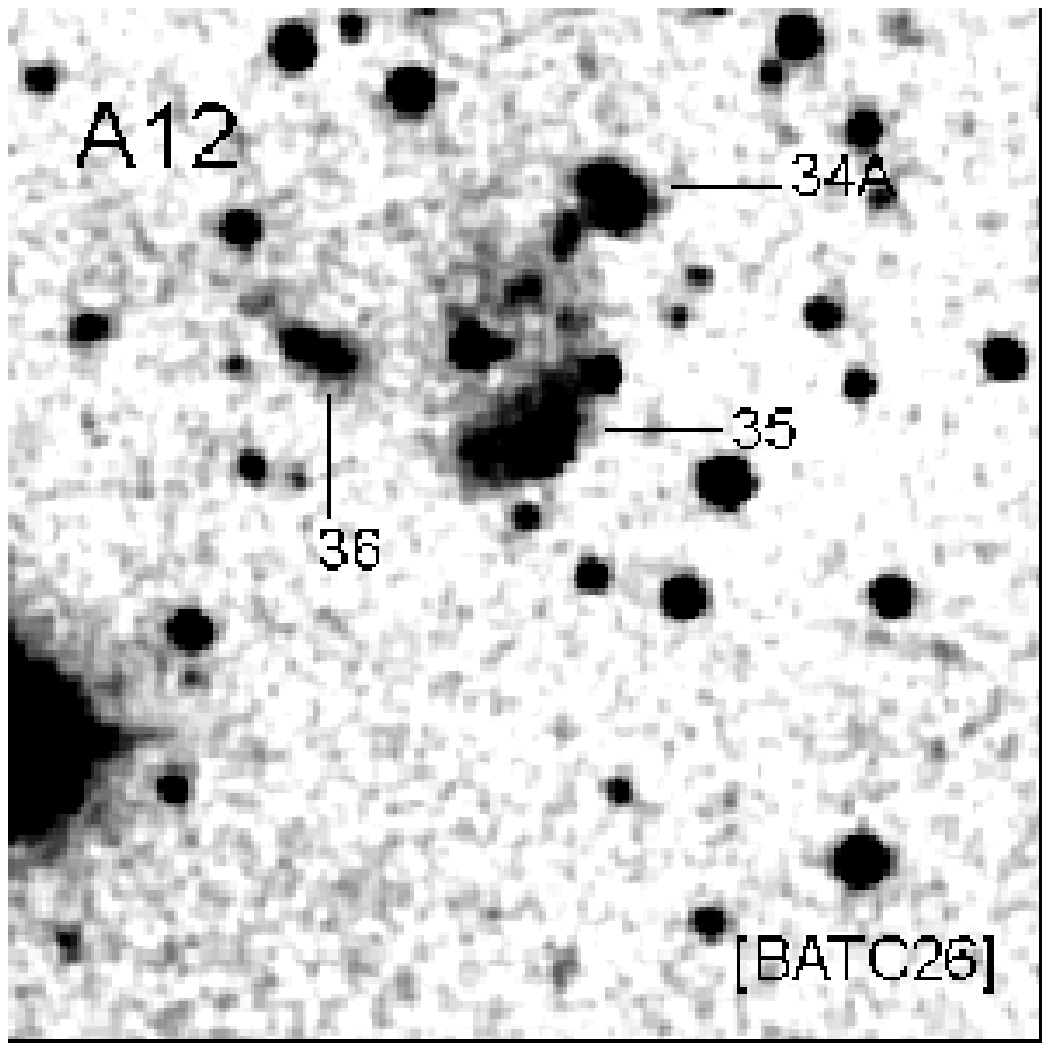,height=5.3cm,width=5.3cm}\hskip 1mm\psfig{figure=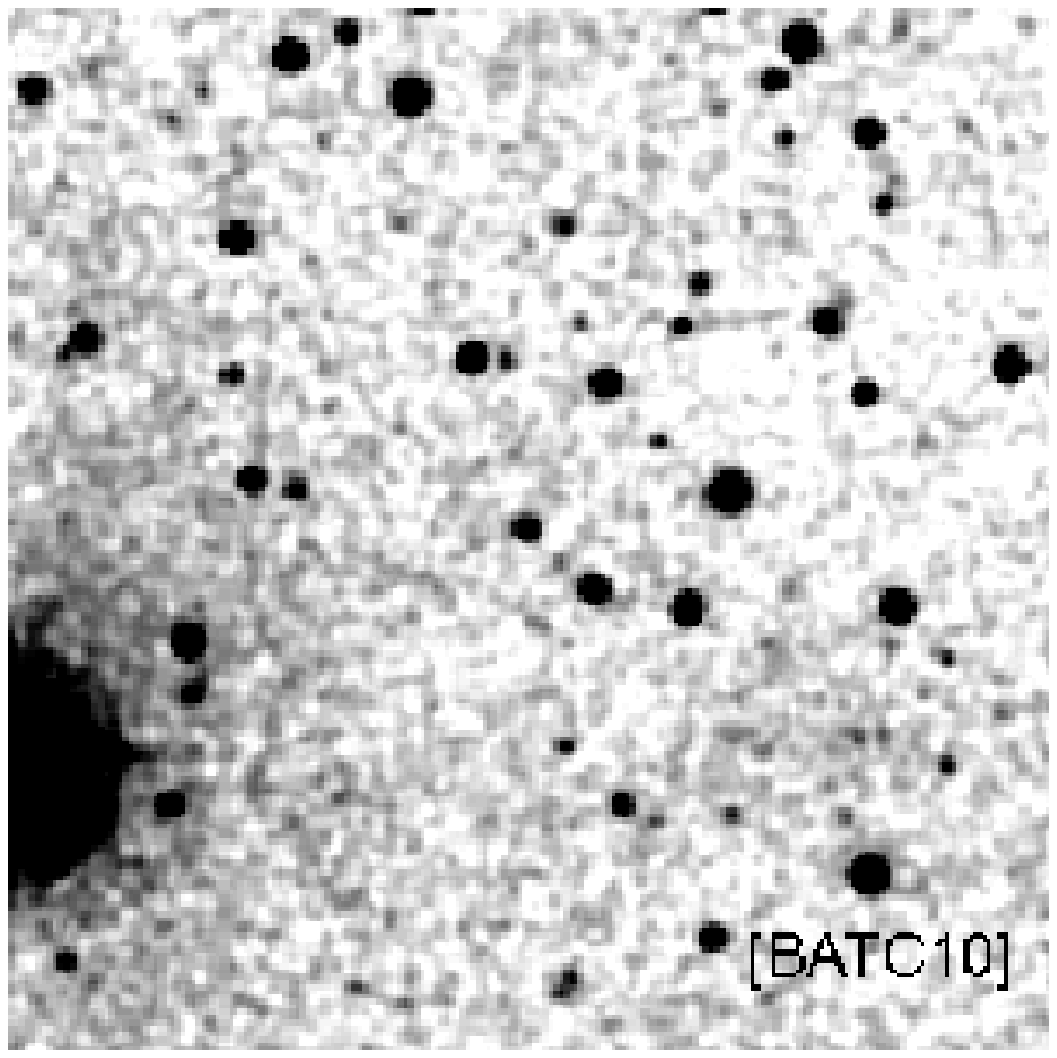,height=5.3cm,width=5.3cm}\hskip 1mm\psfig{figure=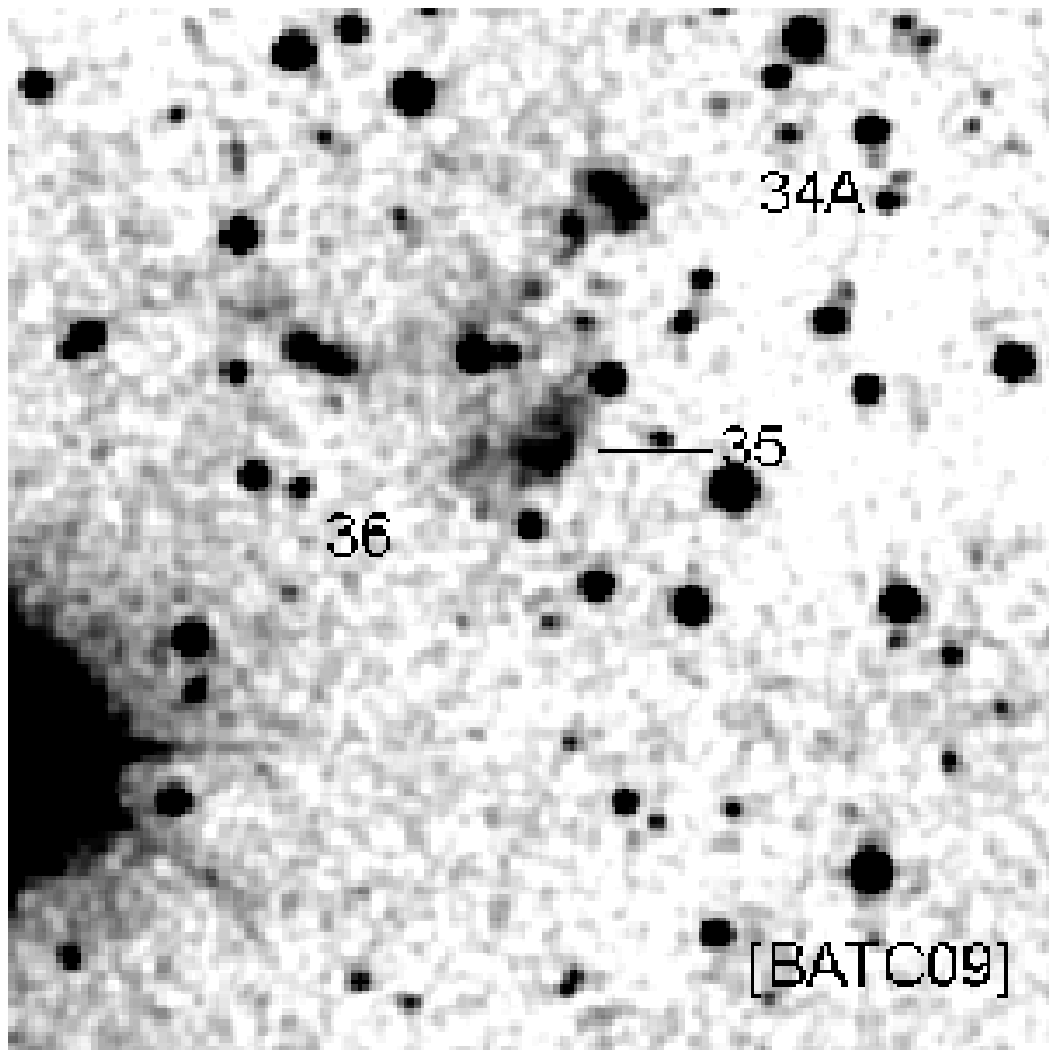,height=5.3cm,width=5.3cm}}

\caption{The [BATC26], [BATC10], and [BATC09] band images of 4 sample fields of our surveys, each row corresponds to a field. For details see text.}\label{fig1}
\end{figure}

In order to demonstrate the effectiveness of our surveys, we present in Figure~\ref{fig1} four sample fields out of the long list of 
Table 3. For each field, we give 3 images taken with our filter set in a row of
the image array of Figure~\ref{fig1}. The first
column is the narrow band [BATC26] images, the second and third columns are [BATC10] and [BATC09] band images, respectively. The HH objects detected in both [BATC26] and [BATC09] bands are markeded using the numbers assigned in our surveys. In all the fields, 
HH objects show up only in [BATC09] and [BATC26] bands, with
their images more prominent in [BATC26], whereas they are completely 
invisible in [BATC09] band.
The first field, labeled {\it A02} in the top row, has dimensions of
512$\times$512 pixels, in which the HH objects 1A, 1B, 1C and 4, 5 are coincident, respectively, with
HH279C, HH279B, HH279A and HH317, HH318 in Bally et al.'s list
(Bally et al. 1996, 1997), therefore this field may also serve as a comparison of our observations with theirs. Comparing with their work, it is clear that all the HH objects
falling into the region overlaped with Bally et al.'s (1997)
were picked up using our method.
The rest rows are 3 other HH object survey fields: A06, A10, and A12, whose sizes are, respectively, 200$\times$200, 300$\times$300 and 
200$\times$200 in pixels.
A02 is located in L1455, A12 is in the vicinity of CI Tau, 
and A06 is located at about 1\arcdeg east of NGC1333.

\section{Results and Discussion}

From the 56 target fields, we  discovered more than 150 HH candidates by comparison
between [BATC09] and [BATC10] frames. From the 15 [BATC26] fields, we identified
68 new HH objects.  These objects are listed in  
Table 3. In the table, column 1 is the numbers of the newly detected HH objects assigned in our surveys,
columns 2 and 3 are, respectively, the right ascension and declination in 
1950 epoch. The source positions were derived using the Guide Star Catalog. 
The overall accuracy is estimated to be about 0.5 arcsec (Fan et al. 1996).
A breaf comment to each object is given in column 4.  Among the 68 objects,
15 objects coincide with those reported by Bally, Devine, \& Reipurth (1996) and 
Bally et al. (1997).  These objects are all in the fields A02 and A05 , 
which are the central
part of NGC~1333 in Perseus.

\par

According to their morphological properties, the new HH objects are classified
into knots, jets, arcs, patches, and cirri.  
Most of the 68 new HH objects are either bright or faint knots,
while a few of them are cirri or cirrus groups, arcs, and jets.  There are
7 HH objects showing arc structures, including No.1A, 7, 8, 11, 60, 62 and 63.
 Objects No.1, 9, 31, 34, and 56 are remarkable in that they show extended jets.
HH No.1A, which is
coincident with HH 279C in Bally et al. (1997), extends to No.1C (HH279A
in Bally et al. 1997) forming a large jet structure,
as clearly shown in the top row of Fig~\ref{fig1}.
HH objects No.31A and 31B are two main knots of a spectacular jet in the 
field A10, which is located at about 2\arcdeg east of Barnard 5. 
It is also possible that the 
jet-like HH flows of HH No.31A, 31B, and 31C are the northern part of a bow structure with the apex at HH No.31A. 
We note that about one third of the newly discovered
HH objects occur in groups. By the term ``occur in group'' we mean that in the vicinity of 0.5 pc from a HH object there is at least one other
HH object.

\par

For the local star forming regions, where the fields are rarely contaminated
by the compact HII regions, We believe that the filter combination of [BATC09] and [BATC10] is an efficient and reliable procedure to find new HH objects.

\begin{table}[t]
\caption{A List of New Herbig-Haro Objects}\label{table3}
{\tiny
\begin{tabular}{cccl}
\hline\hline
PMO&RA.&DEC.&Comments\\
No&(1950)&(1950)&\\
\hline
 1A &3:24:14.56 &30:06:50.5  &arc (HH279C)*\\
 1B &3:23:54.97 & 30:08:13.7 & jet (HH279B)*\\
 1C &3:23:53.59 & 30:09:09.4 & cirrus (HH279A)*\\ 
 2  &3:23:55.03 & 31:25:41.2 & bright knot with star\\     
 3  &3:23:55.42 & 30:15:31.1 & knot (HH278)*\\  
 4  &3:24:07.40 & 30:01:29.1 & cirrus (HH317)*\\  
 5  &3:24:40.65 & 30:03:53.3 & knot (HH318)*\\  
 6A &3:25:07.45 & 31:09:21.8 & knot (HH338A)*\\
 6B &3:25:06.12 & 31:07:52.6 & cirrus (HH338C)*\\
 7  &3:25:10.03 & 30:39:36.3 & arc (HH351)*\\
 8  &3:25:44.91 & 30:59:17.6 & arc (HH341)*\\
 9A &3:25:48.99 & 30:55:02.2 & knot (HH343A)*\\
 9B &3:25:46.55 & 30:55:23.5 & jet (HH343D)*\\
10  &3:25:49.30 & 30:42:06.9 & two knots  \\
11  &3:25:49.44 & 30:54:24.0 & arc (HH350)*\\
12A &3:25:55.40 & 31:02:55.9 & knot (HH344A)*\\
12B &3:25:53.98 & 31:01:57.1 & knot (HH344B)*\\
13A &3:26:10.15 & 31:05:04.9 & knot (HH347B)*\\
13B &3:26:11.59 & 31:05:07.4 & knot (HH347A)*\\
14  &3:26:12.64 & 30:49:33.6 & knot (HH352)*\\
15  &3:26:21.76 & 31:03:16.6 & cirrus (HH348)*\\
16A &3:26:32.79 & 31:20:02.5 & patch (HH353)*\\
16B &3:26:28.14 & 31:19:56.1 & cirrus\\
17  &3:26:30.68 & 31:03:17.2 & cirrus (HH349)*\\
18  &3:27:28.59 & 30:17:36.7 & comma-like knot\\
19  &3:27:33.11 & 30:11:43.4 & knot\\
20  &3:27:40.89 & 30:19:11.1 & knot \\
21  &3:27:42.39 & 30:27:50.9 & knot\\
22  &3:27:48.56 & 30:14:28.1 & faint patch\\ 
23A &3:27:52.26 & 30:24:40.0 & faint patch \\
23B &3:27:55.29 & 30:25:07.6 & faint patch\\
24  &3:28:42.35 & 30:59:50.3 & bright knot\\
25  &3:28:45.84 & 30:59:43.5 & bright knot\\
26  &3:29:21.71 & 31:14:31.8 & faint patch \\
27  &3:29:58.81 & 31:16:12.7 & bright knot with faint cirrus\\
28  &3:30:09.16 & 30:59:03.5 & knot\\
29A &3:30:27.49 & 30:59:40.8 & knot\\
29B &3:30:29.77 & 30:58:47.3 & group of  knots\\
30A &3:30:43.72 & 30:54:54.5 & knot with cirrus\\
30B &3:30:44.37 & 30:54:58.3 & knot\\
30C &3:30:44.48 & 30:55:01.3 & knot\\
30D &3:30:45.03 & 30:55:20.1 & knot\\
31A &3:43:49.77 & 32:36:03.8 & bright knot,SE part of the jet\\
31B &3:43:53.81 & 32:37:06.9 & knot,middle part of the jet\\
31C &3:43:58.57 & 32:37:58.8 & patch,NW part of the jet\\
32  & 3:43:56.54&  32:33:47.2&  several faint knots\\
33  & 4:30:12.46&  22:48:51.8&  knot \\
34A & 4:31:12.99&  23:03:16.2&  bright knot\\ 
34B & 4:31:14.22&  23:03:09.4&  knot,middle part of jet\\      
34C & 4:31:15.26&  23:02:47.1&  knot,middle part of jet\\
35  & 4:31:14.90&  23:01:55.7&  bright knot\\
\hline
\end{tabular}
}
\end{table}

\setcounter{table}{2}

\begin{table}[t]
\caption{-Continued}\label{table 3}
{\tiny
\begin{tabular}{cccl}
\hline\hline
No&RA.&DEC.&Comments\\
&(1950)&(1950)&\\
\hline

36  & 4:31:20.15&  23:02:28.0&  bright knot  \\
37A & 5:39:14.36&  -1:46:28.1&  bright knot\\
37B & 5:39:11.50&  -1:47:52.8&  bright knot \\
38  & 5:44:02.11&   0:24:59.0&  knot with cirrus\\ 
39  & 5:44:07.94&   0:24:34.2&  knot\\
40  & 5:44:31.73&   0:20:45.7&  bright knot\\
41A & 5:44:33.49&   0:10:45.5&  knot\\
41B & 5:44:36.90&   0:09:58.4&  knot\\
41C & 5:44:38.39&   0:09:59.5&  knot\\
42  & 5:44:33.96&   0:23:51.8&  faint knot\\
43A & 5:45:33.12&   0:17:58.3&  knot\\
43B & 5:45:30.00&   0:18:14.7&  diffuse faint patch\\
44A & 5:45:45.02&   0:24:46.3&  knot\\
44B & 5:45:44.13&   0:24:39.4&  faint knot\\
45  & 5:47:25.85&   2:52:14.9&  2' diameter complex\\
46  & 5:48:31.27&   2:41:06.9&  knot\\
47  & 5:49:26.63&   3:01:33.0&  knot\\
48  & 5:50:55.10&   2:40:58.7&  knot \\
49  & 5:51:33.01&   2:36:37.7&  knot\\
50A & 6:04:27.85&  -5:53:02.0&  faint knot\\ 
50B & 6:04:27.28&  -5:54:05.1&  faint knot \\
51A & 6:13:03.15&  14:18:51.8&  bright arc-like knot\\
51B & 6:13:05.50&  14:18:22.7&  knot \\
51C & 6:13:07.32&  14:17:25.6&  knot\\
51D & 6:13:07.91&  14:17:56.3&  knot\\
52A & 6:37:02.24&  10:08:50.9&  knot\\
52B & 6:37:00.58&  10:09:21.6&  curved chain\\
52C & 6:36:53.47&  10:09:38.7&  cirrus\\
52D & 6:36:50.59&  10:10:01.6&  cirrus\\
53  & 6:37:33.83&  10:21:36.2&  bright complex\\
54  & 6:37:45.51&   9:56:16.5&  knot with diffuse patches\\
55  & 6:37:46.25&  10:10:47.0&  knot\\
56  & 6:37:48.39&  10:12:46.5&  chained knots or jet\\
57A & 6:37:50.44&  10:42:42.3&  bright knot\\
57B & 6:38:02.18&  10:41:26.3&  bright patch\\
57C & 6:38:13.99&  10:38:53.9&  patch\\
57D & 6:38:24.57&  10:38:39.7&  patch\\
57E & 6:38:37.43&  10:42:27.9&  knot with cirrus\\
58A & 6:37:54.97&  10:14:57.6&  knot alined with 58B,58C\\
58B & 6:37:55.64&  10:15:17.2&  knot \\
58C & 6:37:55.85&  10:15:36.5&  knot\\
59  & 6:37:59.95&   9:35:42.4&  knot\\
60  & 6:38:01.08&  10:08:10.4&  arc\\
61A & 6:38:09.89&  10:09:34.1&  bright knot\\
61B & 6:38:08.54&  10:09:11.7&  bright knot\\
62  & 6:38:13.75&  10:16:41.1&  5' bright arc\\ 
63  & 6:38:13.76&   9:35:48.8&  arc \\
64A & 6:38:27.30&   9:32:16.1&  bright patch\\ 
64B & 6:38:31.04&   9:32:03.1&  bright patch\\
65  & 6:38:37.15&   9:33:34.2&  bright patch \\
66  & 6:38:38.54&   9:30:55.5&  bright knot \\
67  & 6:38:42.36&  10:26:41.3&  1' diameter complex\\ 
68  & 6:57:11.21&  -4:46:54.2&  two knots\\
\hline
\multicolumn{4}{l}{(*) Detected also by Bally, et al. (1996, 1997)}\\
\end{tabular}
}
\end{table}

\acknowledgments
We specially thank the staff of BATC Beijing group for observation time
allocation, for their helpful discussion, for their hospitality and excellent
support. This research is supported in part by NSFC grant 19673020.

\end{document}